\documentclass[prl,twocolumn,superscriptaddress,nofootinbib]{revtex4-1}
\usepackage{graphicx}
\usepackage{bm}
\usepackage{amssymb}
\usepackage{color}
\usepackage{amsmath}
\usepackage{mathrsfs}
\usepackage{amstext}
\usepackage[english]{babel}
\usepackage{latexsym}
\usepackage{array}             
\usepackage{multirow}         
\usepackage[usenames,dvipsnames]{xcolor}
\usepackage[colorlinks=true,citecolor=Blue,linkcolor=RubineRed,urlcolor=Blue]{hyperref}
\usepackage{tocloft}
\usepackage{multibib}

\def\la{\langle}
\def\ra{\rangle}

\newcommand{\beq}{\begin{equation}}
\newcommand{\eeq}{\end{equation}}
\newcommand{\beqa}{\begin{eqnarray}}
\newcommand{\eeqa}{\end{eqnarray}}

\begin{document}
\title{Universal breakdown of  Kibble-Zurek scaling  in  fast quenches across a phase transition}
\author{Hua-Bi Zeng} 
\author{Chuan-Yin Xia}
\affiliation{ Center for Gravitation and Cosmology, College of Physical Science
and Technology, Yangzhou University, Yangzhou 225009, China}
\author{Adolfo del Campo}
\affiliation{Department  of  Physics  and  Materials  Science,  University  of  Luxembourg,  L-1511  Luxembourg, Luxembourg}
\affiliation{Donostia International Physics Center,  E-20018 San Sebasti\'an, Spain}

\begin{abstract}
The crossing of a continuous phase transition gives rise to the formation of topological defects described by the Kibble-Zurek mechanism (KZM) in the limit of slow quenches. The KZM predicts a universal power-law scaling of the defect density as a function of the quench time.
We focus on the deviations from KZM  experimentally observed
in rapid quenches and establish their universality.
While KZM scaling holds below a critical quench rate, for faster quenches the defect density and the freeze-out time become independent of the quench rate and exhibit a universal power-law scaling with the final value of the control parameter. These predictions are verified in several paradigmatic scenarios in both the classical and quantum domains.
\end{abstract}

\maketitle

The Kibble-Zurek mechanism (KZM)
is one of the cornerstones in nonequilibrium physics, with applications ranging from condensed matter to quantum computing
\cite{kibble1976,kibble1980,Zurek1985,Zurek1996,Dziarmaga10,Polkovnikoc2011,Campo2014}.
It predicts and explains universal scaling laws in the dynamics across a phase transition driven by modulating a control parameter $\lambda$ across a critical point $\lambda_c$ in a finite quench time scale $\tau_Q$.
The KZM exploits the  divergence of the equilibrium relaxation time
$\tau=\tau_0/|\epsilon|^{\nu z}$
as a function of the proximity to the  critical point
\begin{equation}
\epsilon=(\lambda_c-\lambda)/\lambda_c,
\end{equation}
$\tau_0$ is a microscopic constant.
Critical slowing down is thus
characterized by the correlation-length critical exponent $\nu$ and the dynamic critical exponent $z$. This motivates the use of the adiabatic-impulse approximation \cite{Zurek1996,DamskiZurek06,Dziarmaga10,Campo2014}.
According to it, far from the critical point, both in the high and low symmetry phases, the relaxation time is small and the system adjusts to a variation of the control parameter $\lambda(t)$ instantaneously. By contrast, near the critical point $\lambda_c$ the system is effectively frozen.
The transition from the frozen to the adiabatic stages is estimated by the KZM to occur at the freeze-out time scale, identified by matching the time elapsed after crossing the critical point to the instantaneous relaxation time,  $\hat t=\tau(\lambda(\hat{t}))$.
Linearizing the quench around the critical point  so that $\lambda(t)= \lambda_c(1-t/\tau_Q)$ and $\epsilon(t)=t/\tau_Q$, one finds that the freeze-out time scales as 
\begin{equation}
\hat{t} \sim (\tau_0 \tau_{Q}^{z \nu})^{\frac{1}{1+z \nu}}.
\label{Eq2}
\end{equation}
As the order parameter begins to grow significantly,
a mosaic of domains form. According to the KZM, the average domain size $\hat{\xi}$  is set by the equilibrium correlation length of the order parameter
at $\lambda(\hat{t})$, i.e., $\hat{\xi}=\xi(\lambda(\hat{t}))$.  At the interface between adjacent domains, topological defects are generated. Consider point-like topological defects such as kinks and vortices.
The equilibrium scaling law for the correlation length reads
$\xi(\lambda) = \xi_0 |\epsilon|^{-\nu}$, where $\xi_0$ is a microscopic constant. Evaluating it at the freeze-out time $\hat{t}$,  the average domain size  is set by $\hat{\xi}=\xi_0(\tau_Q/\tau_0)^{\frac{\nu}{1+z\nu}}$ and determines  the average defect density $n$ after the phase transition \begin{equation}
n \sim \frac{1}{\xi(\lambda(\hat{t}))^d} \propto \tau_{Q}^{-\frac{d \nu}{1+z\nu}},
\label{EqnKZM}
\end{equation}
{where $d$ is the spatial dimension}.
This universal scaling law as a function of the quench rate is the key prediction of the KZM. It has been experimentally
tested in several systems, such as liquid crystals  \cite{Chuang1991,Bowick1994,Digal1999},
superfluid helium \cite{Hendry1994,Bauerle1996,Ruutu1996,Dodd1998}, Josephson junctions  \cite{Carmi2000,Monaco2002,Monaco2003,Monaco2006}, thin film superconductors \cite{Maniv2003,Golubchik2010},  a linear optical
quantum simulator \cite{Xu2014}, trapped ions \cite{Pyka2013,Ulm2013,Ejtemaee2013},  {quantum annealers \cite{Bando2020,King2022}, } and ultracold gases \cite{Sadler2006,Weiler2008,Lamporesi2013,Navon2015,Donadello2016,Ko2019}.
It was recently proposed \cite{Campo2018,Ruiz2020} that the
fluctuations of $n$ are universal and described by a distribution in which the mean, the variance, and all other cumulants inherit the power-law scaling with the quench time predicted by the KZM. These universal features beyond the scope of the conventional  KZM have  been observed in simulations  \cite{Campo2018,Ruiz2020,Xia2020,Mayo21,Campo2021,Subires21,Li202101,Li202102} and confirmed in experiments \cite{Cui2020,Goo2021,Bando20,King2022}.

In addition to the verification of the  KZM scaling, deviations from the KZM predictions
have also been reported. These are relevant to applications involving quenches at fast and moderate rates.
Numerical simulations indicate the breakdown of the KZM scaling laws with the onset of a plateau in which the defect density is independent of the quench rate, e.g., { in confined ion chains}
 \cite{delcampo10,DeChiara10}, a holographic superconducting ring \cite{Sonner2015}, and a one-dimensional quantum ferromagnet \cite{GomezRuiz19}. As long as the quench time is shorter than the
timescale in which the order parameter grows, the defect formation dynamics is insensitive to the quench rates and yields
a constant defect density \cite{Chesler2015}.
Experimentally, deviations from KZM have been observed in ultracold Bose and Fermi gases driven through the normal-to-superfluid phase transition by a rapid quench  \cite{Donadello2016,Ko2019,Goo2021,Shin2022c}.

One may argue that even point-like topological defects are characterized by a finite healing length. Intuitively, there is a maximum number of defects that a system of finite size can support. However, deviations from KZM occur at densities well below those at maximum packing.
A plausible explanation of the plateau relies on the fast relaxation of defects at high densities via the annihilation of pairs with opposite topological charge (e.g., vortices and antivortices in a superfluid)
and the presence of coarsening \cite{Chandran12,Chesler2015,Libal20,Mukhopadhyay20,Das21}. In experiments with superfluid gases, it is challenging to
detect individual vortices in a turbulent condensate after a fast quench \cite{Donadello2016,Ko2019,delcampo10,Liu2018v1,Goo2021}. Using two-rate driving schemes, first proposed in \cite{Antunes06}, the onset of the plateau in recent Bose gas experiments \cite{Goo2021} has been attributed to  early-time coarsening before the freeze-out time scale \cite{Shin2022c}.


A quantitative understanding of fast quenches is currently lacking.
What is the exact mechanism leading to the emergence of the plateau in the defect density? At what quench rates do the KZM scaling laws break down?
How does the plateau value of the defect density depend on the depth of the quench? Are any of these features universal?
We introduce an extension of KZM  that addresses all of these questions.

\begin{figure}[t]
\includegraphics[width=0.9\linewidth]{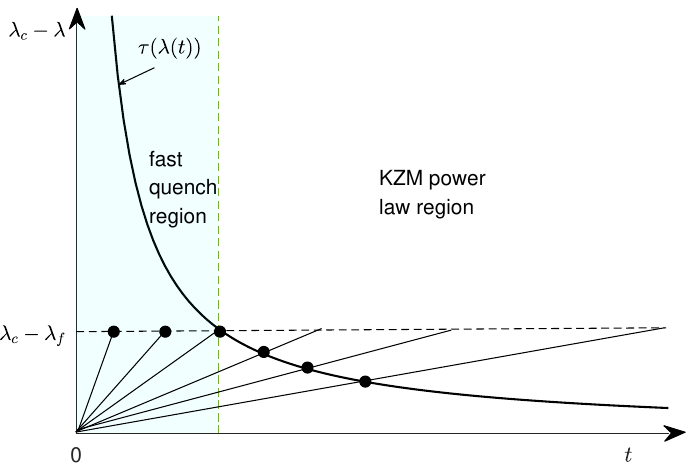}
\caption{Universal breakdown of KZM at fast quenches. The KZM relies on the adiabatic impulse approximation to determine the freeze-out time $\hat{t}$,  by equating the time elapsed after crossing the phase transition to relaxation time. For slow quenches, $\hat{t}$ is given by the power law (\ref{Eq2}). For fast quenches of finite depth $\lambda_f$,  it is set by the equilibrium relaxation time with $\hat{t}\sim \tau(\lambda_f)$, and becomes independent of the quench time. Thus,  the density of defects saturates at a plateau, leading to the breakdown of the KZM scaling relations at fast quenches.} \label{Fig1}
\end{figure}

\textit{Universal deviations from KZM in a fast quench.---}
The KZM identifies  the time scale $\hat{t}$
in which the system begins to partition into independent
domains, whose average size is set by the nonequilibrium correlation length
 $\hat{\xi}$.
 Heuristically, KZM assumes that the instantaneous relaxation time in the broken symmetry phase becomes arbitrarily small as the quench progresses. Building on this assumption and the adiabatic-impulse approximation, the freeze-out time $\hat{t}$ in Eq. (\ref{Eq2}) and all other nonequilibrium quantities, such as $\hat{\xi}$ and $n$ in (\ref{EqnKZM}), inherit a universal scaling with the quench time $\tau_Q$. However, realistic quenches in numerical simulations and experiments terminate at a finite value of the control parameter $\lambda_f$, which sets a lower limit to the relaxation time in the broken symmetry phase, see Fig. \ref{Fig1}.
For quenches of finite depth, we propose that the freeze-out time is set by
\begin{equation}
\hat{t}\sim{\rm max}(\tau(\lambda(\hat{t})),\tau(\lambda_f)).
\end{equation}
For rapid quenches,
the system  is characterized by a single value of the freeze-out time
\begin{equation}
\hat{t} \sim \tau(\lambda_f) \propto \epsilon_{f}^{-z\nu },
\label{Eqthat}
\end{equation}
 independent of the quench time $\tau_Q$, where  $\epsilon_{f}= (\lambda_c-\lambda)/\lambda_c$. As a result, in the spirit of KZM, the
average domain size  for different quench rates  is set by the equilibrium correlation length $\hat{\xi}=\xi(\lambda_f)$,
which naturally explains the plateau of defect density  that appears in the limit of rapid quenches.
This allows us to predict the relationship between defect density and the final
value of the control parameter $\lambda_f$ setting the plateau value
\begin{equation}
n \sim \frac{1}{\xi(\lambda_f)^d} \propto  \epsilon_{f}^{d \nu},
\label{Eqnplateau}
\end{equation}
that is universal and independent of the quench time; see also \cite{Chesler2015} for a related result in holographic systems.
We define the first critical quench rate $\tau_Q^{c1}$ by equating
the time at which the quench ends at $\lambda_f$,   $t_f=\tau_Q^{c1} (\lambda_c-\lambda_f)/\lambda_c$,  to the relaxation time at $\lambda_f$. The condition
\begin{equation}
\tau_Q^{c1} \left(1-\frac{\lambda_f}{\lambda_c}\right)=\frac{\tau_0}{|\epsilon_f|^{z\nu }}
\label{Eq6}
\end{equation}
 yields the expression for the  critical quench rate
\begin{equation}
\tau_Q^{c1} \propto   \epsilon_{f}^{-(z\nu +1)}.
\label{Eq7}
\end{equation}

\begin{figure}[t]
\includegraphics[width=1\linewidth]{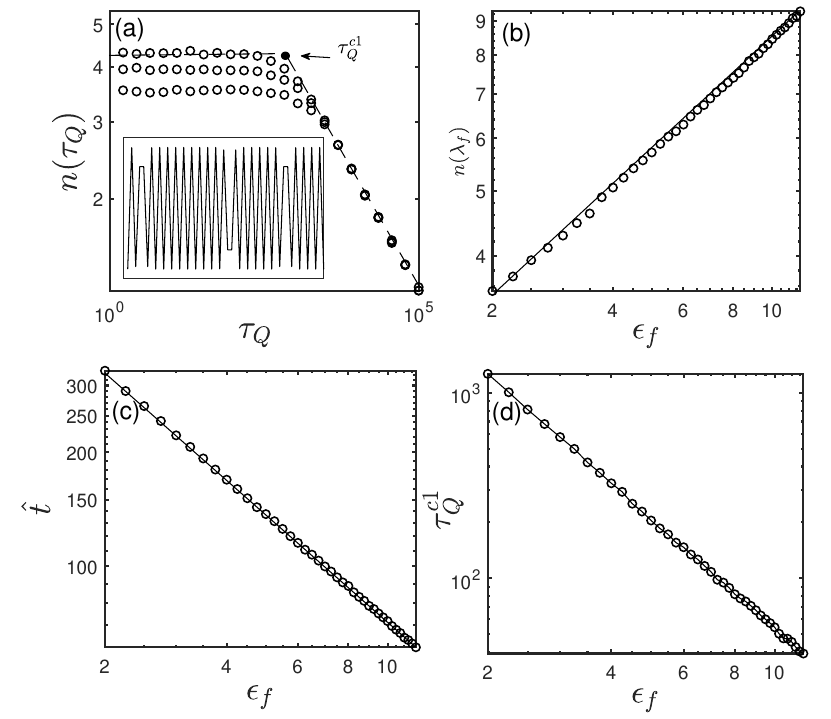}
\caption{Universal breakdown of KZM at fast quenches in a one-dimensional $\phi^4$-model in Log-Log plots. (a) The dependence of the density on the quench time is shown for three different values of $\lambda_f=-2,-1.5,-1$, from top to bottom. The inset shows a single realization of the scalar field $\phi_n$, in the zigzag phase with excitations in the form kinks (at the interface of adjacent
zigzag domains) after crossing the phase transition.  In the KZM scaling regime, $n=(21.1 \pm 0.4) \tau_Q^{-0.245 \pm 0.007}$, while the onset and value of the plateau depend on $\lambda_f$.
The value of the density (b) and the freeze-out-time (c) at the plateau, as well as the first critical quench rate $\tau_Q^{c1}$ (d), exhibit a universal power-law scaling with the quench depth. The plateau data is fitted by $n(\lambda_f)=(2.53 \pm 0.04) \epsilon_f^{0.494 \pm 0.005}$, $\hat{t}=(628\pm 11)\epsilon_f^{-0.95 \pm 0.03}$ and $\tau_Q^{c1}=(5022 \pm 57) \epsilon_f^{-1.99 \pm 0.02}$.
}\label{Fig2}
\end{figure}

According to the picture (\ref{Fig1}), the defect density is a constant when the quench time
is shorter than $\tau_Q^{c1}$, and abruptly turns into the KZM power-law for $\tau_Q>\tau_Q^{c1}$.
The critical quench rate $\tau_Q^{c1}$ can also be defined as the intersection
point of the plateau and the KZM power law. Indeed, by equating Eqs. (\ref{EqnKZM}) and (\ref{Eqnplateau}), Eq. (\ref{Eq7}) is  recovered.
We distinguish $\tau_Q^{c1}$ from a well-understood second critical quench time associated with the onset of adiabaticity when the nonequilibrium correlation length is comparable to system size, $\xi(\hat{t})\sim L$, in the limit of slow quenches; see e.g.,  \cite{Dziarmaga05}.
We have thus identified universal deviations of the KZM at fast quenches. The quantities $\hat{t}$ and $n$   exhibit a universal scaling relation with the depth of the quench and are independent of the quench rate. We note that this dependence is not a feature predicted by the conventional KZM but may be present in the scaling regime \cite{delcampo10,DeChiara10,delcampo11}.
In what follows, we verify these predictions in three paradigmatic models of phase transitions in one and two spatial dimensions.

\textit{Fast dynamics of spontaneous parity breaking.--- }
Consider a lattice version of a one-dimensional $\phi^4$ theory describing a second-order phase transition that involves spontaneous parity breaking, a canonical testbed for  KZM scaling \cite{Laguna97,Laguna98,Ruiz2020}.
The equilibrium critical exponents take  the mean-field values  $\nu=1/2$ and $z=2$. 
This model is of relevance to structural phase transitions between a linear phase and a  zigzag phase, exhibited by confined ion chains \cite{delcampo10,DeChiara10}, Wigner crystals, and colloids. It involves a real scalar field $\phi_l(t)$ with $l=1,\dots,L$, and the global potential
\begin{equation}
\label{poteq}
V(\{\phi_l\},t)=\sum_{l=1}^L\frac{1}{2}[\lambda(t)\phi_l^2+\phi_l^4]+c\sum_{l=1}^{L-1}\phi_l\phi_{l+1},
\end{equation}
where $\lambda(t)$ is the  control parameter.
With coupling constant $c=1/2$, the critical point is $\lambda_c= 2c$, and  in the high-symmetry linear phase, where $\lambda>\lambda_c$, the order-parameter vanishes, i.e., $\la \phi_l\ra=0$ for all $l$; see e.g. \cite{Laguna97,Ruiz2020}. By contrast, the case $\lambda<\lambda_c$ describes a doubly-degenerate broken-symmetry zigzag phase. Varying $\lambda(t)$ across the critical point gives rise to the formation of $\mathbb{Z}_2$ kinks.  Consider the linear ramp $\lambda(t)= \lambda_c(1-t/\tau_Q)$. The time evolution is described by the  coupled Langevin equations
\begin{equation}
\ddot{\phi}_l+\eta \dot{\phi}_l+\partial_{\phi_l}V(\{\phi_i\},t)+\zeta_l=0,
\end{equation}
and is characterized by a dissipation strength $\eta$ and noise fluctuations described by real Gaussian processes fulfilling $\overline{\zeta_l(t)}=0$ and $\overline{\zeta_l(t)\zeta_m(t')}= 2 \eta T\delta_{lm}\delta(t-t')$. Numerical integration is used to generate an ensemble of 50000 trajectories in which to study the density of kinks and its dependence on $\tau_Q$, choosing  $L=100$, $\eta=50$ and $T=2 \times 10^{-5}$. A typical single realization of the order parameter after crossing the phase transition is shown in the inset of Fig. \ref{Fig2}(a), where the kinks separate adjacent zigzag domains.
Figure \ref{Fig2} further shows the breakdown of the KZM scaling law (\ref{EqnKZM}) at fast quenches. The density of defects at the plateau exhibits a universal power-law scaling with the depth of the quench $\epsilon_f$ in agreement with the universal prediction $n \propto \epsilon_f^{1/2}$(\ref{Eqnplateau}).
Figure \ref{Fig2} also shows that the corresponding freeze-out time scales universally at fast quenches according to (\ref{Eqthat}) with the quench depth. Specifically, $\hat{t} \propto \epsilon_f^{-1}$
where the $\hat{t}$ is identified in the simulations as the time when the order parameter begins to grow rapidly \cite{Das2012,Sonner2015}.

{\textit{Rapid quenches in a quantum Ising chain.---}
Let us next explore the universal behavior with fast quenches in  the quantum realm, choosing as a testbed  the quantum 
Ising chain,
\begin{equation}
H=-J \sum_{l=1}^L \left[g(t)\sigma_l^x + \sigma_l^z \sigma_{l+1}^z\right],
\end{equation}
 used to test the KZM in a  quantum phase transition theoretically \cite{Jacek2005,Zurek2005}, as well as in recent experiments using programmable quantum annealers \cite{Bando2020,King2022}.
Here $\sigma_l^{x},~\sigma_l^{z}$ are Pauli operators, $J>0$ is the Ising ferromagnetic coupling,
and $g(t)$ acts as a magnetic field that favors spin alignment
along the $x$-axis. We assume periodic boundary conditions $\vec{\sigma}_{N+1}=\vec{\sigma}_{1}$ and consider a linear quench $g(t)=g_i-t/\tau_Q$ across $g_c=1$ to drive the  quantum phase transition
from  the paramagnetic phase ($g_i>g_c$) to a ferromagnet.
The breaking of parity symmetry gives rise to the formation of $\mathbb{Z}_2$ kinks \cite{GarciaPintosTielas19}, but under unitary dynamics, these are described by coherent quantum excitations \cite{Jacek2005,Zurek2005,Dziarmaga2012}.}

{By adopting the Jordan-Wigner transformation, the model can be  rewritten in terms of spinless
fermions, and the dynamics described by time-dependent
Bogoliubov-de Gennes equations \cite{Nishimoribook2011,Jacek2005}. According to the KZM,   kinks appear in pairs under periodic boundary conditions. We focus on the ferromagnetic case, where the number of kink pairs is described by the operator
\begin{equation}
\hat{P}=\frac{1}{4}\sum_{j=1}^{L}\big(1-\sigma_j^z \sigma_{j+1}^z\big),
\end{equation}
where $L$ is the total number of sites.
The expectation value of this operator in the nonequilibrium state resulting from the crossing of the quantum phase transition exhibits the KZM scaling law $P=\langle \hat{P}\rangle \propto \tau_Q^{-1/2}$, with the equilibrium critical exponents $z=1, \nu=1$ \cite{Zurek2005}.
Following \cite{Jacek2005,Zurek2005}, we confirm  the scaling law and report the onset of a plateau at fast quenches by numerical solving the dynamical
Bogoliubov-de Gennes equation, by quenching the system from a paramagnetic phase to an anti-ferromagnet state, 
with size $L=1000$ and other parameters  chosen to be $J=1$, $g_i=2$ and $g_f=0$, see Fig. \ref{Fig3} (a).  The predicted fast-quench scaling $P/L \propto  \epsilon_{f}^{d \nu}$ is also supported as in Fig.  \ref{Fig3} (b).
Thus, our findings also hold in the quantum regime as we have verified in the 1D transverse field Ising model.
\begin{figure}[t]
\includegraphics[width=1.0\linewidth]{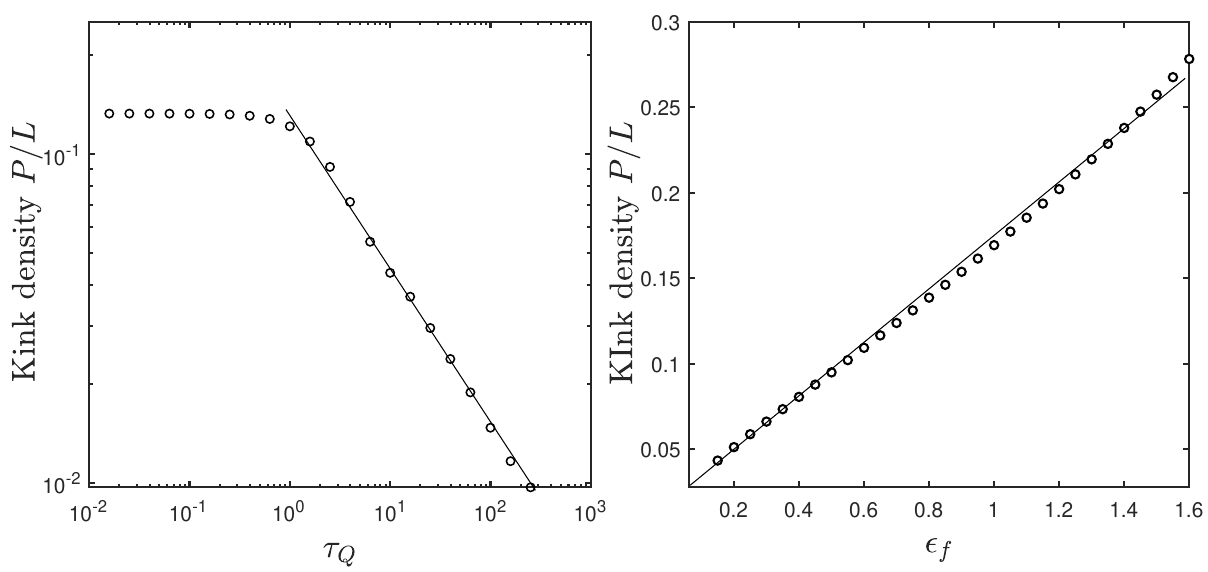}
\caption{Universal deviations from the KZM in the 1D transversal field Ising model after a fast thermal quench. (a) The KZM power-law scaling of the kink
density is interrupted by the onset of a plateau as the quench time is reduced. A fit to the data reads $P/L=(0.14 \pm 0.03) \tau_Q^{ -0.54 \pm 0.05}$ in the KZM regime.  (b)
In the fast quench case, the kink density is independent of the quench rate and scales universally with the quench depth $\epsilon_f$, with data fitted to $P/L=(0.177 \pm 0.004)\epsilon_f^{0.99 \pm 0.05}$.}\label{Fig3}
\end{figure}
}

{
\textit{Vortex density saturation in two spatial dimensions.---}
To verify the universality of the fast-quench dynamics in two spatial dimensions, we analyze the spontaneous vortex formation in a newborn scalar superfluid, associated with the spontaneous breaking of $U(1)$ symmetry. To this end, we use a mean-field description supplemented by thermal fluctuations, that seed symmetry breaking. Specifically, we consider the stochastic Gross-Pitaevskii equation \cite{Cockburn}, previously used to test KZM  \cite{DamskiZurek10,Sabbatini11,Das2012,SuGou13,Liu20}.
  In two spatial dimensions, it reads}
\begin{equation}
(i-\gamma)\frac{\partial\phi}{\partial t} =
-\frac{1}{2}\left(\frac{\partial^{2}\phi}{\partial x^{2}}+\frac{\partial^{2}\phi}{\partial y^{2}}\right)
+\lambda(t)\phi + \tilde{g}|\phi|^{2}\phi + {\eta(\vec{x},t),}
\label{SGPE}
\end{equation}
\noindent
{where
$\phi = |\phi(\vec{x})|e^{i\theta(\vec{x})}$ is the condensate wavefunction, $\gamma$ represents the dissipation rate, $\tilde{g}$ sets the strength of the the non-linearity, and
$\eta(\vec{x},t)$ is the thermal noise
satisfying the fluctuation-dissipation relation
$\langle \eta(\vec{x},t)\eta^{\ast}(\vec{x}^{\prime},t^{\prime})\rangle
= 2 \gamma T\delta(\vec{x}-\vec{x}^{\prime})\delta(t-t^{\prime})$.
{The control parameter inducing the transition is the chemical potential $-\lambda(t)$. The critical point is located at
$\lambda=0$, below which $\phi$
acquires a finite value, in agreement with the  Ginzburg-Landau
theory of continuous phase transition.
The critical exponents
take the mean-field values  $\nu=1/2$, $z=2$.
We  induce the phase transition by quenching $\lambda$ as
\begin{equation}
\lambda(t) = -\frac{t}{\tau_{Q}},
\label{mu}
\end{equation}
\noindent
from an initial value $ \lambda=0$ to a finite final one
$\lambda_f < 0$, and allow the system enough time
to thermalize initially and stabilize eventually from initial vanishing $\phi$ with small fluctuation $\langle \delta \phi(\vec{x})\delta \phi^{\ast}(\vec{x}^{\prime})\rangle
= 10^{-6} \delta(\vec{x}-\vec{x}^{\prime})$. As long as a very small $T$ is chosen, the initial conditions are important and the rapid quench limit could be
expected to be largely independent of the noise.
Numerically, 100 trajectories are generated by integrating  (\ref{SGPE}), with a  system size $L=100$  in the spatial directions $x$ and $y$, and using $316$ Fourier modes. In the time direction,  the fourth-order Runge-Kutta (RK4) method is used with a time step $\Delta t=0.025$.
The newborn 2D superfluid is proliferated by vortices as shown in the inset of Fig. \ref{Fig4}. Choosing $T=2\times10^{-3}$, $\tilde{g}=1$ and $\gamma=1$,}  the universal scaling laws governing the plateau of the vortex density at fast quenches are characterized as a function of the quench depth in Fig. \ref{Fig4}. Panels (a) and (b) report the scaling behavior of the vortex density and freeze-out time matching
 the predictions $n \propto  \epsilon_{f}^{d \nu}$ and $\hat{t} \propto \epsilon_{f}^{-\nu z}$ when $d=2, \nu=1/2, z=2$ accurately. Thus, this numerical example confirms the universality of the fast-quench critical dynamics in two spatial dimensions. Notice that the KZM predicted $n(\tau_Q) \propto \tau_Q{1/4}$ with fixed $T_f$ is found (not shown here), which matches the observation in \cite{Sabbatini11,Das2012,SuGou13,Liu20}.

\begin{figure}[t]
\includegraphics[width=1.0\linewidth]{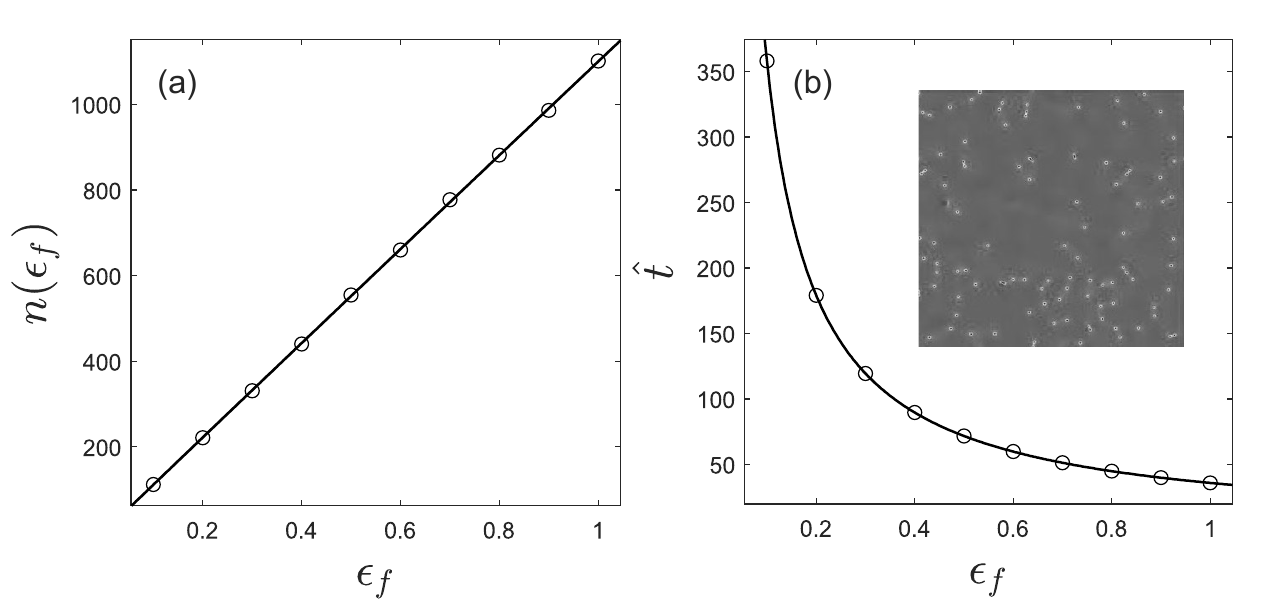}
\caption{Universal deviations from the KZM in a 2D superfluid after a fast thermal quench. (a) The vortex number is independent of the quench rate and scales universally with the quench depth $\epsilon_f$. A fit to the data reads $n_{\lambda_f}=(1100 \pm 5) \epsilon_f^{0.99 \pm 0.01}$.  (b) Corresponding universal scaling of the freeze-out time as function of $\epsilon_f$, with data fitted to $\hat{t}=(35.9 \pm 0.2)\epsilon_f^{-0.99 \pm 0.03}$. The inset shows the distribution of vortices in a single realization.}\label{Fig4}
\end{figure}

\textit{Discussion and summary.---}{We have established the universality of deviations from the KZM occurring at fast quenches across a continuous  phase transition, observed in recent experiments.
As a result of these deviations, the power-law scaling predicted by the KZM  for the defect density as a function of the quench rate is interrupted by the onset of a plateau for quenches exceeding a critical quench rate. The plateau defect density, associated freeze-out time, and the critical quench rate are found to scale with the amplitude of the quench following universal power laws. We have confirmed these predictions in  a classical lattice $\phi^4$ model of relevance to ion chains, as well as a one-dimensional quantum Ising chain, thus establishing the universality of fast critical dynamics in the quantum domain. In addition, we have  verified the universal scaling at fast quenches in two spatial dimensions by characterizing the saturation of the vortex density in a newborn superfluid.}

{Our results broaden the application of equilibrium scaling theory to nonequilibrium phenomena in the limit of fast quenches, without restrictions to slow driving or adiabatic perturbation theory. The scaling laws predicted here are directly testable in any platform previously used to explore KZM in either the classical or quantum regimes, such as trapped ions \cite{Pyka2013,Ulm2013,Ejtemaee2013,Cui2020}, colloids \cite{Keim15}, ultracold gases \cite{Sadler2006,Weiler2008,Lamporesi2013,Navon2015,Donadello2016,Ko2019}, multiferroics \cite{Griffin12,Lin2014}, Rydberg quantum simulators \cite{Keesling2019}, and annealing devices \cite{Bando2020,King2022}.}

{\it Acknowledgements.---}
This work is supported by the National Natural
Science Foundation of China (under Grant No. 12275233).

\bibliography{rapidKZMbib}
\end{document}